\documentclass[11pt,twoside]{article}

\usepackage{graphicx}
\usepackage{asp2006}
\usepackage{epsf}
\usepackage{psfig}
\usepackage{lscape}

\begin{document}
\title{Dust in the Early (z$>$1) Universe}   
\author{Fabian Walter}   
\affil{Max Planck Institut f\"ur Astronomie Heidelberg, Germany}    

\begin{abstract} 
  Although dust emission at cosmological distances has only been
  detected a little more than a decade ago, remarkable progress has
  been achieved since then in characterizing the far--infrared
  emission of high--redshift systems. The mere fact that dust can be
  detected in galaxies at high redshift is remarkable for two reasons:
  (a) even at very early cosmic epochs (all the way to the first Gyr
  of the universe), dust production was apparently very effective, (b)
  due to the inverse K--correction (`the magic of (sub--)millimeter')
  is it actually possible to detect this dust emission using current
  facilities.  Deep blind surveys using bolometer cameras on single
  dish telescopes have uncovered a population of massively starforming
  systems at z$\sim$2, the so--called submillimeter galaxies (SMGs).
  Follow--up radio and millimeter interferometric observations helped
  to characterize their main physical properties (such as
  far--infrared luminosities and implied star formation rates).
  Average FIR properties of fainter optically/NIR--selected classes of
  galaxies have been constrained using stacking techniques.  Targeted
  observations of the rare quasars have provided evidence for major
  star formation activity in quasar host galaxies throughout cosmic
  times.  Molecular gas and PAH features have been detected in both
  SMGs and quasars, providing additional evidence for major star
  formation episodes (SFR$\sim$500-3000\,M$_{\odot}$\,yr$^{-1}$) in
  the brightest systems.  Even though remarkable progress has been
  achieved in recent years, current facilities fail to uncover the
  counterparts of even major local starbursts (such as Arp\,220) at
  any significant redshift (z$>$0.5). Only ALMA will be able to go
  beyond the tip of the iceberg to study the dust and FIR properties
  of typical star forming systems, all the way out to the epoch of
  cosmic reionization (z$\gg$6).

\end{abstract}


\section{Some Thoughts/Disclaimer}

This conference (Cosmic Dust -- Near and Far) covered an impressive
range of topics related to dust in the universe. Dust emission from
evolved stars and supernovae, protoplanetary and debris disks, as well
as our solar system was discussed in detail. One session was dedicated
to the global dust emission in our Galaxy as well as in other galaxies
--- and a particular emphasis was put on recent laboratory studies of
dust and the modeling of dust emission and evolution. Progress in this
field of research has clearly been dramatic in the last decade!

I was asked to review the subject of `Dust at high redshifts', a
subject that has exploded in recent years: only a decade ago, hardly
any observations of high--redshift dust existed. In the following I
will attempt to summarize key aspects of dust at high redshift and
discuss some of the recent results --- this review cannot be complete
and the reader may forgive me for not giving credit to everybody's
work that has been published on this subject. In this review, I will
try to address a mainly `Galactic' audience/reader as most of the
presentations at the conference were centred around Galactic dust
emission. For further reading, the interested reader is also referred
to some excellent reviews in the literature (e.g., Blain et al.\ 2002,
Ivison 2001, Smail et al.\ 2002, 2003, Solomon \& Vanden Bout 2005,
Smail 2006).

\begin{figure}[b!]
\begin{center}
\includegraphics[height=.3\textheight,angle=0]{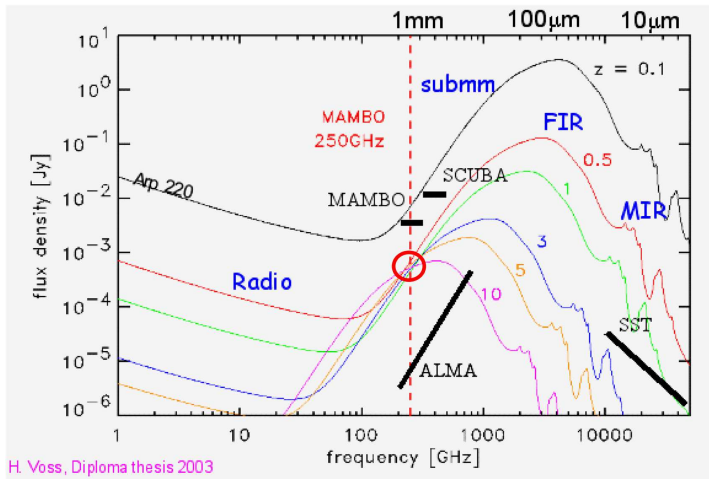}
\end{center}
\caption{The observed spectral energy distribution of Arp\,220 as a
  function of redshift. The horizontal dashed line indicates the
  observing wavelength at $\sim$1\,mm (e.g., using the MAMBO
  bolometer). The observed flux density at this frequency is almost
  independent of redshift for 0.5$<$z$<$10 (as indicated by the
  circle). This is the reason why high--redshift dust emission can be
  detected given current instrumentation (`magic of
  (sub--)millimeter'). Short horizontal bars indicate $\sim$4 sigma
  sensitivities for MAMBO and SCUBA, respectively. Figure adopted from
  Voss (2003).}
\end{figure}

\section{The Magic of (Sub--)millimeter}

Even with current (sub--)millimeter cameras it is not trivial to map
the dust emission in nearby galaxies (e.g. Guelin et al.\ 1995, Thomas
et al.\ 2002, Meijering et al.\ 2005, Bendo et al.\ 2006, Weiss et
al.\ 2008). How is it then possible that dust can be detected out to
the highest redshifts with the same instruments?  Together with the
fact that some systems at high redshifts are very luminous, the main
reason for this astonishing fact is what people refer to as the `magic
of (sub--)millimeter'. This is illustrated in Figure~1: Here the
spectral energy distribution (SED) of the ultraluminous infrared
galaxy (ULIRG) Arp~220 (a template commonly used in extragalactic
astronomy) is plotted as a function of different redshifts (from z=0.1
to 10). Two things are immediately clear from this plot: (1) The
emission gets fainter and (2) the emission is being redshifted as a
function of redshift. However, if one were to observe such an object
at $\sim1$\,mm wavelengths (indicated by the vertical dashed line for
a hypothetical observation with MAMBO), the observed flux density
stays roughly constant for any redshift in the range 0.5$<$z$<$10!
This is due to the fact that the peak of the SED is shifted towards
the observed frequency -- in other words, the shift towards the peak
of the SED cancels out the dimming due to the increased redshift
(`inverse K correction'). This immediately implies that the flux
density observed of a source at z$\geq$1 in the millimeter regime is
not a function of redshift/distance, it only is a function of the
intrinsic luminosity of the source. This situation is obviously in
stark contrast to what would be seen at any other wavelength
(X--ray/UV/optical/IR/radio), and is, at first glance, contrary to
sanity and reason. It should be noted though that in the example
discussed here (Arp~220) the sensitivities of current bolometers (e.g.
MAMBO) are an order of magnitude too high to detect an Arp~220
counterpart (with an intrinsic FIR luminosity of
5$\times$10$^{12}$\,L$_\odot$) at significant redshifts (see short
horizontal bar in Figure~1). In other words, current instruments only
allow one to observe the brightest objects at any redshift (as
discussed further below). But it is immediately clear that ALMA will
have no problem detecting objects that are more than an order of
magnitude fainter than Arp\,220 (as indicated by the diagonal line in
Figure~1).

\begin{figure}[b!]
\begin{center}
\includegraphics[height=.35\textheight,angle=0]{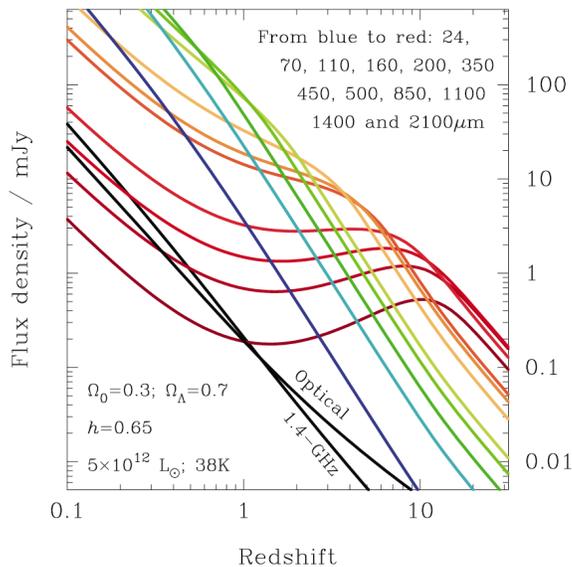}
\end{center}
\caption{A more detailed description of the situation described in
  Figure~1: Here the measured flux density of a source of L$_{\rm
    FIR}$=5$\times$10$^{12}$\,L$_\odot$ is plotted as a function of
  redshift for various observing bands. At around 1\,mm wavelengths
  the measured flux density stays approximately constant as a function
  of redshift. For comparison, the flux densities in the optical and
  radio are also shown. Figure taken from Blain et al.\ (2002).}
\end{figure}

This effect is discussed in more detail in Blain et al.\ (2002):
Figure~2 shows the result of his models: the flux density of a source
with L$_{\rm FIR}$=5$\times$10$^{12}$\,L$_\odot$ is plotted as a
function of redshift for various observing bands. It is clear that
both the optical and radio flux densities drop dramatically as a
function of redshift, whereas the flux densities stay roughly constant
for observations at $\sim$1\,mm. It is this `inverse K correction'
that allows us to observe objects at significant redshifts at all!

This `magic of submillimeter' can be illustrated further by simulated
ALMA deep field observations as shown in Figure~3: In the top panel,
galaxies in an ALMA deep field are separated into low--redshift
(z$<$1.5) and high--redshift (z$>$1.5) objects (left and right panels,
respectively). Due to the reason discussed above, there will be many
more objects that are detectable at high redshifts. This situation is
opposite to what is seen in the optical (bottom two panels, see figure
caption for full credits).

\begin{figure}
\begin{center}
\includegraphics[height=.4\textheight,angle=0]{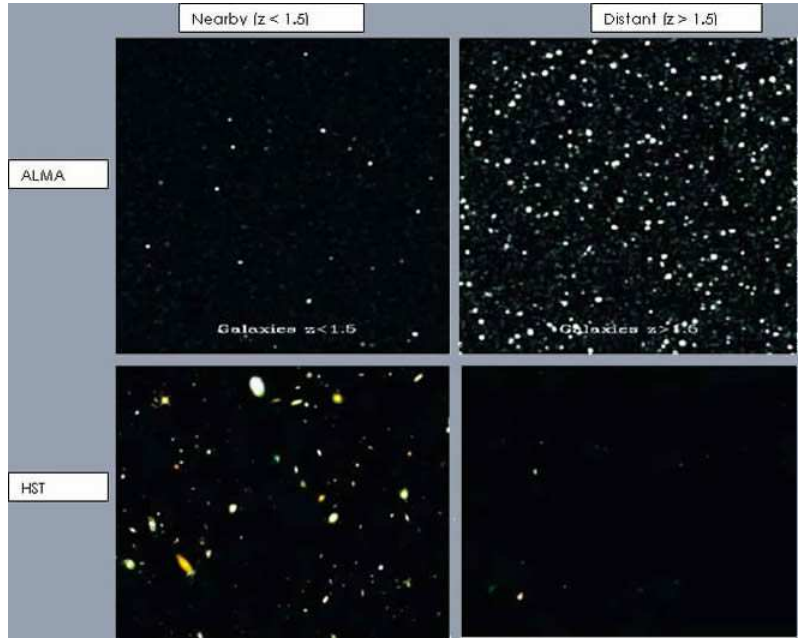}
\end{center}
\caption{Graphical illustration for the fact that the measured flux
  densities at (sub--)millimeter wavelengths at high redshifts are
  roughly constant as a function of redshifts.  Top: Simulation of an
  ALMA deep field, where the left panels shows galaxies at z$<$1.5 and
  the right panel shows galaxies at z$>$1.5.  This relation of number
  densities of detected galaxies is opposite to what is seen in the
  optical (bottom panels).  Top images from Wootten \& Gallimore
  (2000). Bottom images from K.  Lanzetta, K.  Moore, A.
  Fernandez-Soto, and A. Yahil (SUNY) (courtesy of Kenneth M.
  Lanzetta).}
\end{figure}

There are basically two observing strategies for studying the dust
properties of high redshift systems: (A) blind searches using blank
field observations and (B) targeted observations of rare objects. Both
strategies are discussed in the following.

\section{Blank (deep) fields}

The revolution of high--redshift studies of dusty galaxies started in
the late 1990's with the discovery of (sub--)millimeter bright
galaxies (`SMGs') using the SCUBA (JCMT) and MAMBO (IRAM 30m) bolometers
(Smail et al.\ 1997, Bertoldi et al.\, 2000, Smail et al.\ 2002,
Ivison et al.\ 2002, Cowie et al.\ 2002, Dannerbauer et al.\ 2002, and
many more!). One of the most famous studies is the SCUBA deep field
study of the Hubble Deep Field (HDF) by Hughes et al.\ (1998) which
showed that the brightest submillimeter emission in the field did not
show an obvious counterpart in deep HST imaging, as shown in Figure~4
(subsequent studies aimed at identifying a counterpart, Downes et al.\
1999, Dunlop et al.\ 2004).  This difficulty of identifying the
correct optical/NIR counterpart (which can be straightforwardly
explained by the `negative K correction' discussed above if the source
is at high redshift) has been plaguing many of the subsequent
(sub--)millimeter deep field studies: without a secure identification
at optical/NIR wavelengths (or, in fact, at any wavelength),
spectroscopic follow--up using 8--10\,m glass is virtually impossible.
This issue is further complicated by the fact that the beam sizes of
past and current bolometers mounted on single dish telescopes are
large (SCUBA: $\sim$15$''$, MAMBO: $\sim$11$''$), i.e., even in case
there are optical/NIR sources in the immediate neighborhood of a
(sub--)millimeter source, it is often difficult to tell which of the
possible counterparts is in fact the correct one.

\begin{figure}[t!]
\begin{center}
\includegraphics[height=.2\textheight,angle=0]{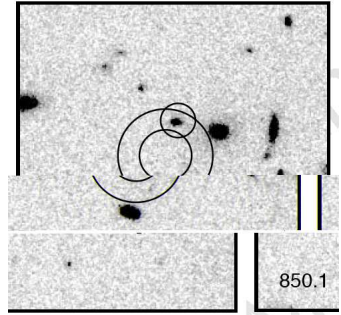}
\end{center}
\caption{The case of HDF850.1, the brightest sub--millimeter source
  discovered by SCUBA in the Hubble deep field. The circle indicates
  the position of the SCUBA source whereas the grey scale represents
  the deep HST imaging. No obvious counterpart can be identified from
  these observations (see text for further details). Figure taken from
  Hughes et al.\ (1998).}
\end{figure}

\begin{figure}[b!]
\begin{center}
\includegraphics[height=.4\textheight,angle=0]{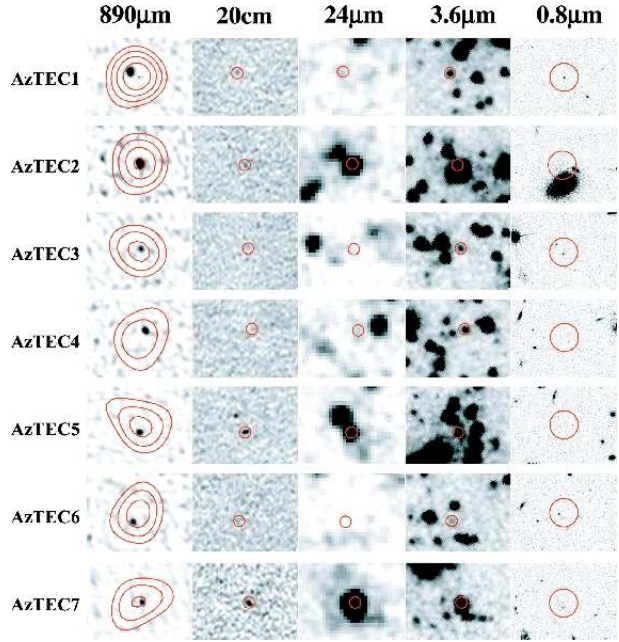}
\end{center}
\caption{Example of how interferometric (sub--)millimeter observations
  can help to pinpoint the emission seen in deep bolometer maps
  (needed to identify the optical/NIR counterpart and for subsequent
  spectroscopic follow--up). The left panels show the AzTEC beam in
  contours overlaid on the SMA detections at 890\,$\mu$m in the COSMOS
  field. Figure taken from Younger et al.\ (2007).}
\end{figure}

Two methods have been pushed forward to address this positional
uncertainty: (A) increase of resolution in the (sub--)millimeter band
using (sub--)millimeter interferometry and (B) increase of resolution
by employing deep radio continuum maps (typically at 20\,cm
wavelengths) -- both methods will be briefly discussed in the
following:

\subsubsection{Millimeter Interferometry:} The most straightforward
method to get better positions of the (sub--)millimeter sources of
interest is by re--observing the candidates with (sub--)millimeter
interferometers. This method works well in practice (e.g., Dannerbauer
et al.\ 2002, Younger et al.\ 2007, Dannerbauer et al.\ 2008),
yielding positional accuracies of $\leq1''$. The major drawback is
that these observations are expensive in terms of telescope time:
typically one full track (i.e., $\sim$8 hours of observing time) is
needed to achieve the required sensitivity at a (sub--)millimeter
interferometer. Also, the primary beam of interferometers at
$\sim$1\,mm wavelengths is small (e.g. in the case of the Plateau de
Bure interferometer: $\sim$20$''$) which implies that only one source
can be followed up at a time. As an example the identification of
AzTEC sources in the COSMOS deep field (Scoville et al.\ 2006, Scott
et al.\ 2008) using SMA interferometry is shown in Figure~5 (Younger
et al.\ 2007) and the identification of one of the brightest SMGs in
the GOODS-North field is presented in Figure~6 (Dannerbauer et al.\
2008).

\begin{figure}
\begin{center}
\includegraphics[height=.15\textheight,angle=0]{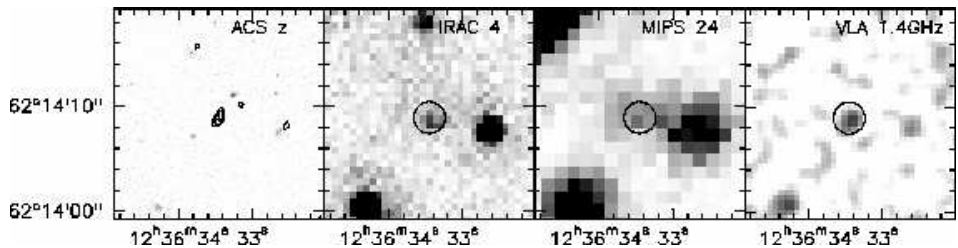}
\end{center}
\caption{Example of how millimeter interferometry was used to pinpoint
  the location of GN10, one of the brightest sources in the GOODS-N
  region. Contours in the left panel indicate the position of the
  source as detected with the IRAM Plateau de Bure interferometer
  (shown as a circle in the subsequent panels). The source is now
  clearly identified, which enables a photometric determination of its
  redshift. Figure taken from Dannerbauer et al.\ (2008).}
\end{figure}

\subsubsection{Radio Interferometry:}

Ivison et al.\ (1998, 2000, 2002) and Smail et al.\ (2000) pioneered a
different technique: identification of SMGs through their radio
continuum emission.  The strength of this approach lies in the fact
that deep ($\sim$10--20$\mu$Jy) wide--field (30$'$) VLA radio
continuum maps exist for most of the deep fields that have been looked
at with (sub--)millimeter bolometers. The advantage of this technique
is that the synthesized radio beam ($\sim$1$''$) is much smaller than
that of single dish bolometers, which in turn enables direct
identification with optical/NIR sources (needed first for photometric
redshift determination and subsequently for spectroscopic follow--up).
The downside of this approach is obviously that sources without a
radio identification can not be identified (the radio completeness in
Ivison et al.\ is $\sim$65$\%$). As the radio flux is a strong
function of redshift (see Figure~2 above) it is likely that the
radio--undetected sources are at high (z$>$4) redshifts (see
discussion below). From the samples of radio--identified SMGs, Chapman
et al.\ (2003, 2005, Figure~7) derived a number of key properties for
this population of sources: the median redshift is z=2.3 ($\sigma$=1.3
-- similar to the median QSO redshift), and the typical FIR luminosity
is $L_{\rm FIR}$=5\,10$^{12}$\,L$_\odot$. Importantly, they conclude
that there is a $>$500 decrease in ULIRG space density from redshifts
z$\sim$2 to 0, implying very strong evolution of this class of
galaxies.

\begin{figure}
\begin{center}
\includegraphics[height=.3\textheight,angle=0]{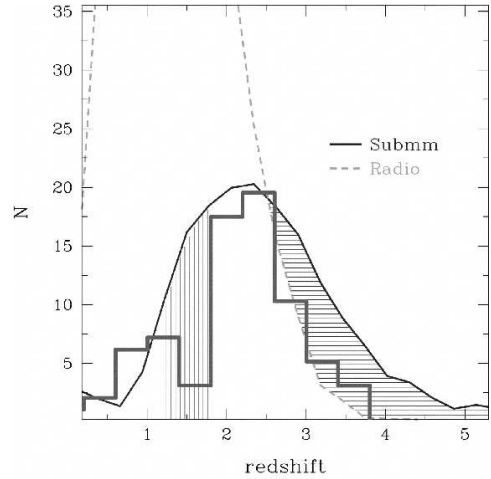}
\end{center}
\caption{Histogram of the redshift distribution of radio--selected
  SMGs. The median redshift of this population is z$\sim$2.3. One of
  the open questions is whether the decrease in SMGs at z$>$4 is an
  intrinsic property of this population, or whether it is due to the
  fact that they are faint in the radio (and thus would not end up in
  the selection shown here). Figure taken from Chapman et al.\
  (2005).}
\end{figure}

By now, 100s of SMGs have been detected in a number of extensive
observing campaigns using SCUBA (e.g.  through the SCUBA Half--Degree
Extragalactic Survey SHADES, Mortier et al.\ 2005, Coppin et al.\
2006, Ivison et al.\ 2007), MAMBO (e.g. Bertoldi et al.\ 2000, 2007),
and AzTEC (e.g., Scott et al.\ 2008, Perera et al.\ 2008). Some of
these surveys were targeting clusters to probe lensed background
sources with the goal to extend the luminosity function of SMGs to
fainter flux levels. The reader is referred to, e.g., Knudsen et al.\
(2006, 2008) and Coppin et al.\ (2007), for a discussion on the faint
end of the SMG luminosity function.

\subsection{Molecular gas and PAH emission in SMGs}

\begin{figure}
\begin{center}
\includegraphics[height=.3\textheight,angle=0]{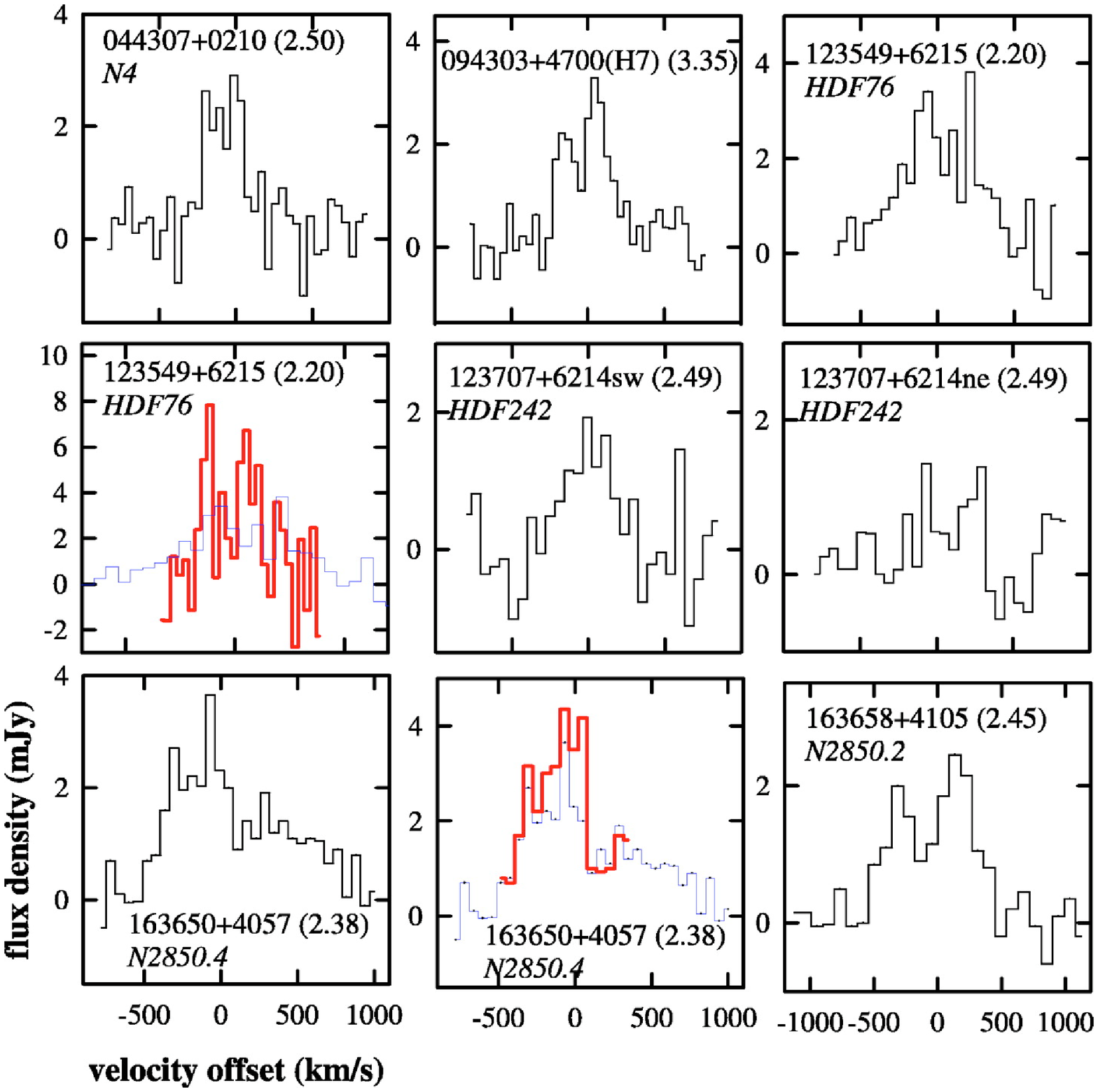}
\end{center}
\caption{Integrated CO spectra of SMGs obtained with the Plateau de
  Bure interferometer. Typically, the CO(3--2) spectrum is shown here;
  the thick lines indicate higher--J CO transitions of two sources.
  Figure taken from Tacconi et al.\ (2006).}
\end{figure}

Both molecular gas and poly--cyclic aromatic hydrocarbons (PAH)
emission (both extensively discussed at this conference) have now been
detected in SMGs. The molecular gas content of SMGs has been studied
in a major observational campaign using the Plateau de Bure
interferometer by Genzel et al.\ (2003), Neri et al.\ (2003), Greve et
al.\ (2005), and Tacconi et al.\ (2006, 2008). Figure~8 shows a composite
of CO spectra for SMGs (Tacconi et al.\ 2006) -- some key results of
this survey are that the implied molecular gas masses are of order
3$\times$10$^{10}$\,M$_\odot$ (an order of magnitude more massive than
the Milky Way, but comparable in mass to the QSO population at similar
redshift). The CO spectra have line widths of
$\sim500\pm200$\,km\,s$^{-1}$, and, together with size measurements,
give dynamical masses of M$_{\rm dyn}\sim10^{11}$\,M$_\odot$, implying
high gas mass fraction of $\sim20-50\%$ (Tacconi et al.\ 2008). In
summary, the detection of abundant molecular gas in these system is
consistent with the picture that they are major starforming systems
that have high FIR luminosities (as evidenced by their strong
sub--millimeter emission).

Emission from PAHs has also been detected in SMGs: Figure~9 shows an
average SMG spectrum at z$\sim$2.5 obtained with the IRS on-board
Spitzer (Valiante et al.\ 2007, see also Lutz et al.\ 2005,
Menendez-Delmestre et al.\ 2007, Pope et al.\ 2008). Both the
6.2$\mu$m and 7.7 $\mu$m PAH features are detected at high
signal--to--noise and the spectrum is well fit by the nearby starburst
M\,82 plus continuum emission. The same study noted that the
PAH/continuum ratio in z$\sim2.5$ SMGs is very similar to what is
found in local ULIRGs. It has also been suggested that the PAH
emission features in SMGs can be used as a proxy for the star
formation rate in SMGs (e.g., Pope et al.\ 2008).

\begin{figure}[b!]
\begin{center}
\includegraphics[height=.3\textheight,angle=0]{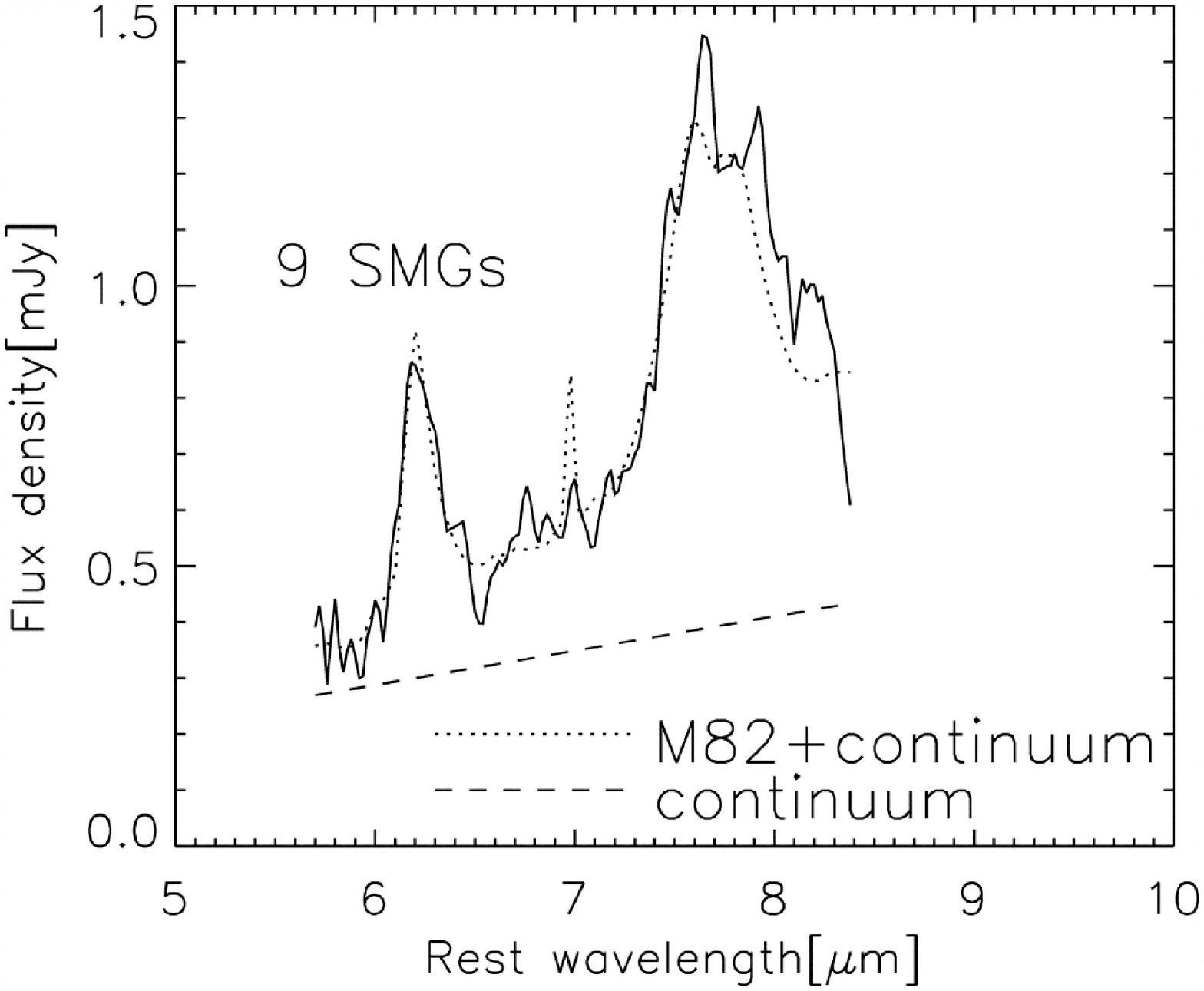}
\end{center}
\caption{Composite Spitzer IRS spectrum for 9 SMGs at a median
  redshift of z$\sim$2.5. The PAH features at 6.2$\mu$m and 7.7 $\mu$m
  are clearly detected and the emission is well fit by an M\,82
  template and an additional continuum component. Figure taken from
  Valiante et al.\ (2007).}
\end{figure}

\subsection{An SMG population at z$>$4?}

One of the open questions is how significant the SMG source population
at z$>$4 really is. Given their high redshifts, it is difficult to
first identify and then spectroscopically confirm these objects.  This
has been pointed out early on by, e.g., Dunlop et al.\ (2001), Eales
et al.\ (2003) and Ivison et al.\ (2005). To date only few SMGs with
confirmed redshifts z$>$4 are known: GN20, GN20.2a and GN 10 (in the
GOODS North field, Daddi et al.\ 2009a, 2009b, Dannerbauer et al.\
2008), and one source in the COSMOS field (Capak et al.\ 2008,
Schinnerer et al.\ 2008). Recently, Coppin et al.\ (2009) identified a
z=4.76 submillimeter--selected source in the Extended Chandra Deep
Field South using the APEX LABOCA survey (Weiss et al., in prep., see
next sub--section), making this the currently highest--redshift source
known that has been detected in a blind survey (Figure~10). All
sources known so far at z$>$4 are potentially very much dust
enshrouded, and it is very likely that ongoing observing campaigns
will soon uncover more systems at similar or even higher redshifts. A
novel technique to select these targets is discussed in Greve et al.\
(2008): at very high redshift (z$>$4), one would expect sources to be
detected by MAMBO but not with SCUBA (given the respective
sensitivities and the slope of the dusty SED, see again Figure~1).
Greve et al.\ identified a number of such `SCUBA drop--out' sources
and argue for either a very high redshift origin of these sources, or
a very cold/different dust composition than typically seen. One should
keep in mind though that the total number of objects that have been
detected in dust emission at z$>$4 is much higher, i.e.  targeted
observations of high--redshift QSOs have resulted in a
(sub--)millimeter detection rate of $\sim1/3$ (as discussed in
Section~4).

\begin{figure}[b!]
\begin{center}
\includegraphics[height=.2\textheight,angle=0]{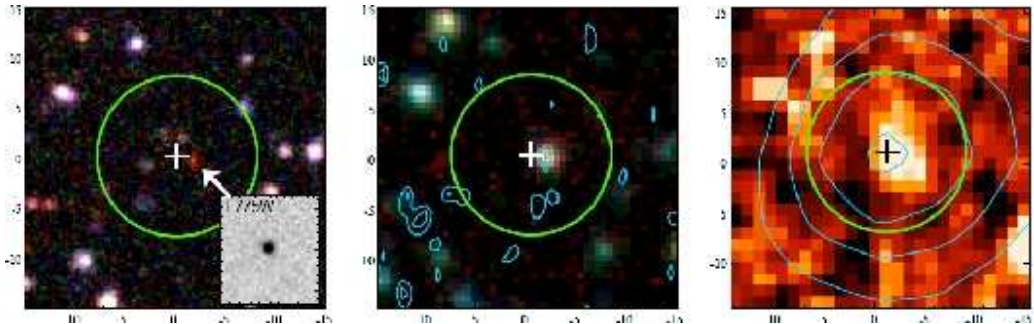}
\end{center}
\caption{Identification of the highest--redshift
  submillimeter--selected source in the Extended Chandra Deep Field
  South using the LABOCA bolometer mounted on APEX. The left label
  shows a BVR image from the MUSYC survey (Gawiser et al.\ 2006), the
  centre is a Spitzer IRAC composite and the right image is a
  24\,$\mu$m image from the Spitzer--Fidel survey (Dickinson et al.,
  in prep.).  The green circle indicates the LABOCA beam, the contours
  in the middle panel the radio continuum emission. All scales are in
  arcseconds. Figure taken from Coppin et al.\ (2009).}
\end{figure}

\subsection{First results from LESS}

The new 870$\mu$m LABOCA submillimeter camera on APEX was recently
used for the LESS project (LESS: `The LABOCA ECDFS Submm Survey').
This survey covers the area of $\sim0.5^\circ\times0.5^\circ$ of the
Extended Chandra Deep Field South (ECDFS), and reaches a uniform rms
of 1.3 mJy over the full area. This is the largest contingent survey
ever performed in the sub--millimeter and the rich auxiliary database
of the ECDFS enables immediate identification of the objects in the
field. Of order 100 objects are individually detected in LESS (Weiss
et al., in prep.), many of which have been identified at other
wavelengths (Coppin et al., in prep.). The auxiliary database also
enables statistical studies of optically/NIR selected galaxies using
stacking techniques. Preliminary results are shown in Figure~11, where
optically/NIR selected galaxies show strong stacked submillimeter
emission (Greve et al.\ 2009). Stacked 870$\mu$m signals are:
0.23$\pm$0.02 mJy (11.5$\sigma$), 0.54$\pm$0.06 mJy (10.8$\sigma$),
0.40$\pm$0.04 mJy (10.0$\sigma$), and 0.46$\pm$0.06 mJy (7.7$\sigma$)
for the K$_{\rm vega}<$20, BzK [B--z,z--K--selection], ERO [extremely
red objects] and DRG [distant red galaxies] samples, respectively (see
Greve et al.\ (2009) for details on the high--redshift source
selection).  Splitting the BzK--selected galaxies up into star-forming
(sBzK) and passive (pBzK) galaxies, the former population is
significantly detected while the latter is not.

\begin{figure}
\begin{center}
\includegraphics[height=.13\textheight,angle=0]{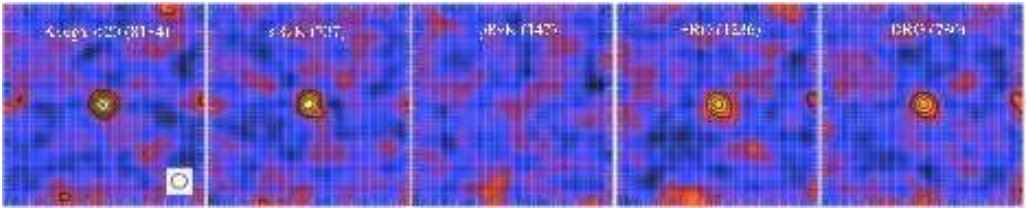}
\end{center}
\caption{First 870$\mu$m stacking results of the LABOCA ECDFS Submm
  Survey (`LESS') of optically/IR selected galaxies. See text for the
  description and flux densities of the respective galaxy class.
  Figure taken from Greve et al. (2009).}
\end{figure}

\section{Targeted Observations of `rare objects'}

The remainder of this review will focus on targeted (sub--)millimeter
observations, in particular quasars. These are so rare on the sky that
chances of detecting them in blind surveys are relatively slim (but
see Coppin et al.\ 2009).

\subsection{General FIR properties of high--z quasars}

Dust has now been detected in many quasars at different redshifts.
This is summarized in Figure~12, where L$_{\rm FIR}$ of the quasars is
plotted as a function of redshift (Beelen et al.\ 2006, Omont et al.\
2003, Wang et al.\ 2007, 2008). Here, full symbols represent
detections, open symbols upper limits, dashed lines indicate the
typical 3 sigma detection limits of MAMBO and SCUBA, respectively.
Quite remarkably, about 1/3 of all sources are detected in the
(sub--)millimeter, independent of redshift. Also, there are hardly any
`outliers' in this plot, indicating that the bright quasars share
similar FIR properties, independent of cosmic age. A major caveat in
this interpretation is the fact that currently only the tip of the
iceberg can be detected with current bolometers -- a situation that is
unlikely to change until the advent of ALMA.

\begin{figure}[b!]
\begin{center}
\includegraphics[height=.3\textheight,angle=0]{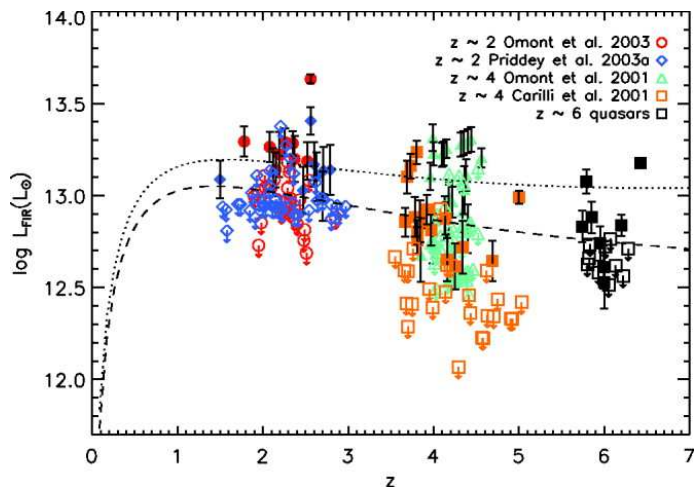}
\end{center}
\caption{L$_{\rm FIR}$ as a function of redshift for all quasars detected so far.  Full symbols  represent detections, open symbols upper limits, dashed lines indicate the typical 3 sigma detection limits of MAMBO and SCUBA, respectively. Figure taken from Wang et al.\ (2007).}
\end{figure}

These individual dust detections can be used to construct a composite
SED of all quasars (each quasar is at a different redshift, thus each
measurement samples a different part of the average [restframe] SED).
The result is shown in Figure~13 (Beelen et al.\ 2006), where the
different measurements have been normalized to the FIR luminosity of
one particular object (see Beelen et al.\ for details). It is striking
how similar the dust SED looks for objects at different redshifts --
amongst other things, this hints at very rapid dust enrichment in the
host galaxies of these objects. In the near future, a guaranteed time
Herschel key project (PI: Klaus Meisenheimer, MPIA) will sample the
dusty SEDs of individual z$>$5 quasars to unprecedented accuracy.

\begin{figure}
\begin{center}
\includegraphics[height=.3\textheight,angle=0]{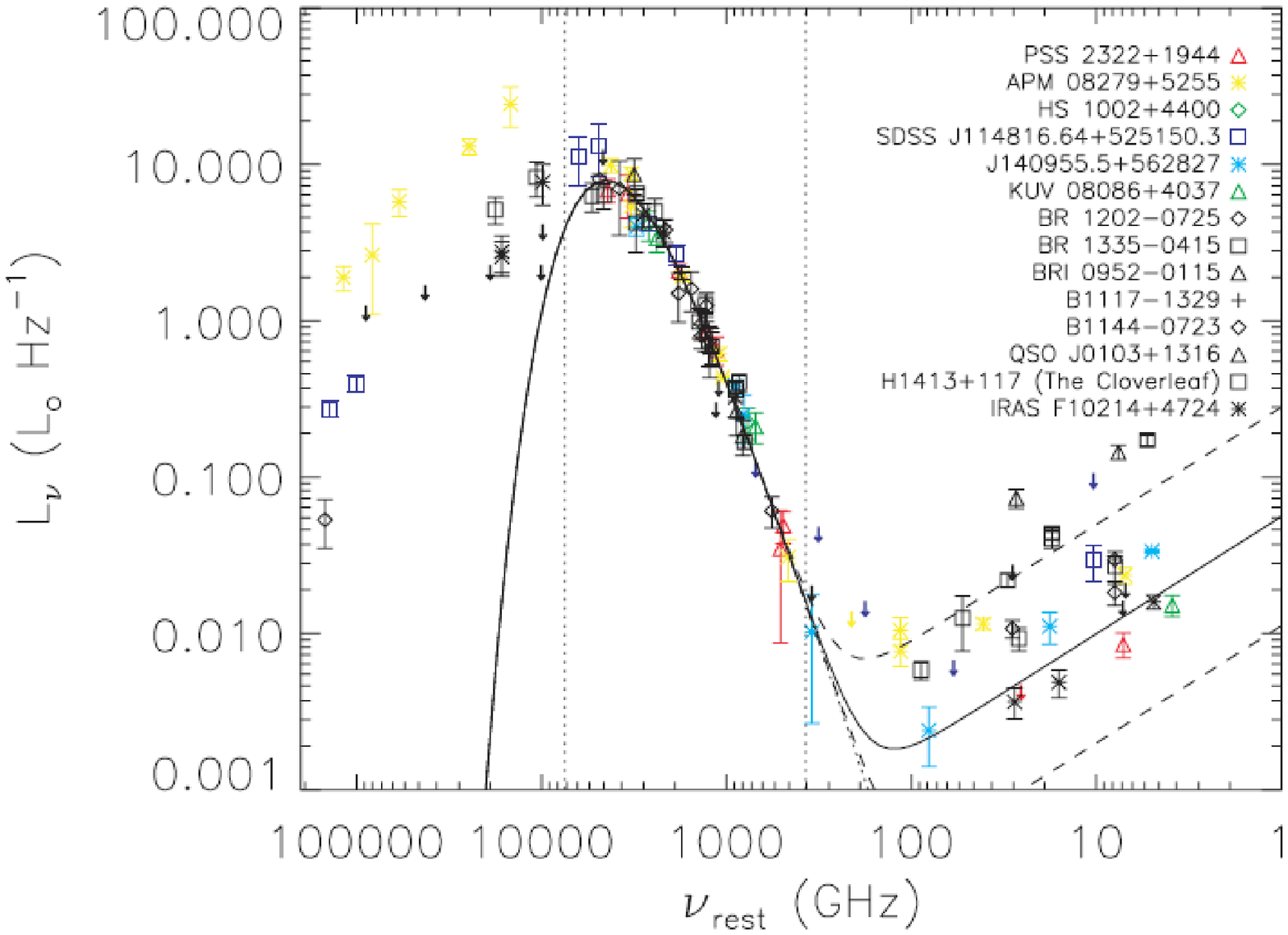}
\end{center}
\caption{Composite high--redshift quasar SED (rest--frame) after
  normalizing the total L$_{\rm FIR}$ emission to the FIR luminosity
  of PSS\,2322+1944. Figure taken from Beelen et al.\ (2006).}
\end{figure}

Another interesting finding is the location of the high--redshift
quasars on the radio--FIR relation (Condon et al.\ 2002, Yun et al.\
2001).  This is shown in Figure~14, where the rest-frame 1.4 GHz
luminosity is plotted as a function of L$_{\rm FIR}$ for different
classes of objects. The crosses show the IRAS 2 Jy sample of Yun et
al.\ (2001); the circles indicate the quasars of Beelen (2003).  The
dashed line shows the mean value of q (quantifying the FIR/radio
relationship, Condon 1992), while the dotted lines display the IR and
radio excesses (5 times above and below the value expected from the
linear far-IR/radio relation seen at low redshift).  The same result
holds if one adds the z$\sim$6 quasar population (Wang et al.\ 2007).
As noted in Beelen et al.\ (2003), the fact that the high-z quasars
roughly follow the Condon relation for star-forming galaxies suggests
that their radio and far--IR emission also arise from star formation.
Some contribution to L$_{\rm FIR}$ by a central AGN can not be
excluded, but it is unlikely that L$_{\rm FIR}$ is predominantly
powered by a central engine.

In this context, it is also interesting to investigate the `star
formation law' in the highest redshifts systems (both quasars and
submillimeter galaxies). In Figure~15, L$_{\rm FIR}$ is plotted as a
function of CO luminosity (a measure for the molecular gas mass). This
plot includes the sample of low-z spiral and starburst galaxies from
Gao \& Solomon (2004), ULIRGs from Solomon et al.\ (1997), z$<$0.2 PG
QSOs from Alloin et al.\ (1992), Evans et al.\ (2001), and Scoville et
al.\ (2003), extrapolated Galactic molecular clouds (GMCs) from Mooney
\& Solomon (1988), the Milky Way (Fixsen et al.\ 1999), and high-z
submillimeter galaxies, radio galaxies, and QSOs from the review by
Solomon \& Vanden Bout (2005). The dashed line is a fit to all sources
and yields a slope of 1.4 (e.g., Riechers et al.\ 2007). The location
of the high--redshift sources (both SMGs and quasars) suggests that
the trend seen at lower redshift (`more SF per unit CO') continues out
to the highest redshifts/luminosities. The particularly interesting
finding is that the quasars and the submillimeter galaxies occupy the
same parameter space (upper right corner in Figure~15). This is
another indication for the fact that the central AGN/quasar does not
contribute significantly to L$_{\rm FIR}$ as one would otherwise
expect the quasar population to be offset from the SMG population in
this plot. It is important to note however that this plot is almost
certainly incomplete, in particular for high luminosities (as both FIR
and CO measurements can only recover the brightest objects given the
sensitivities of current instruments). Indeed, in a recent study Daddi
et al.\ (2008) argue that BzK--selected galaxies fall off this relation
significantly.

\begin{figure}
\begin{center}
\includegraphics[height=.3\textheight,angle=0]{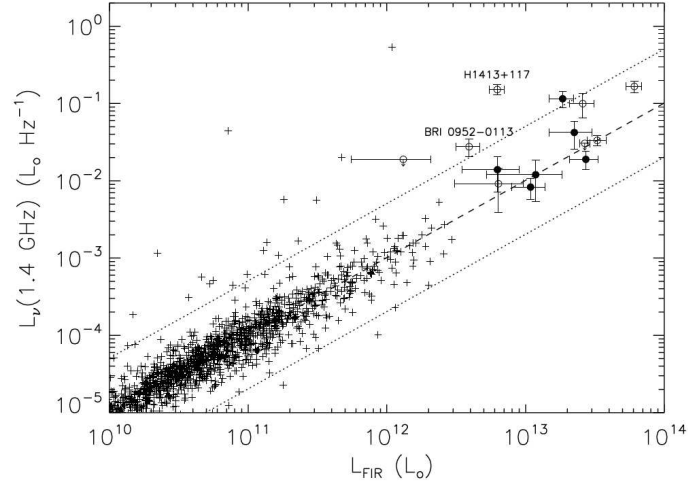}
\end{center}
\caption{The radio--FIR relation for high--redshift quasars. The
crosses show the IRAS 2 Jy sample of Yun et al.\ (2001); the circles
indicate the quasars of Beelen (2003). See text for the description of
the dashed/dotted lines. Figure taken from Beelen et al.\ (2006).}
\end{figure}

\begin{figure}
\begin{center}
\includegraphics[height=.4\textheight,angle=0]{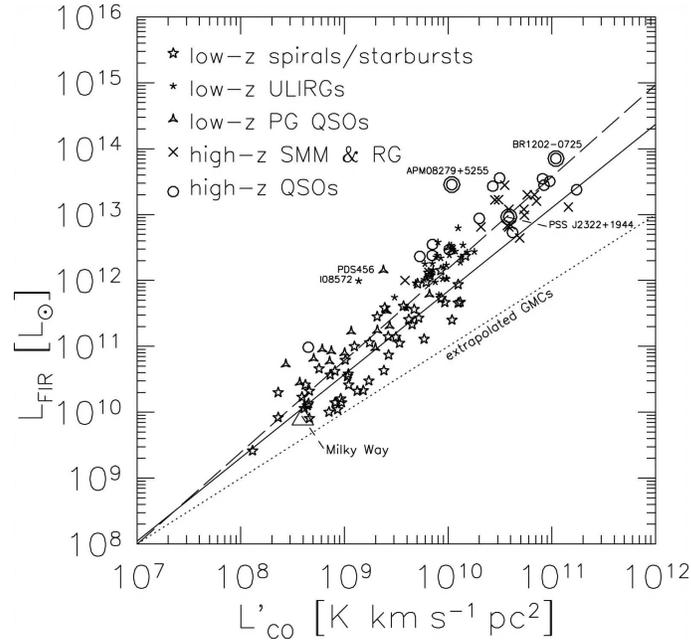}
\end{center}
\caption{The star formation law: L$_{\rm FIR}$ (a measure for the star
  formation activity) as a function of L$'_{\rm CO}$ (a measure of the
  total molecular gas mass). Different galaxy classes are shown here
  (see text for references) -- it is interesting to note that the
  high--redshift SMGs and QSOs cover the same parameter space, hinting
  at only a minor contribution to L$_{\rm FIR}$ by the central
  accreting source in the case of the QSOs. Figure taken from Riechers
  et al.\ (2006).}
\end{figure}

\subsection{Molecular gas and PAHs in QSOs}

We have already seen from Figure~15 that molecular gas is detected in
a number of quasar host galaxies. Like in the case of the
submillimeter galaxies, PAH emission has also been detected in some of
the high--redshift quasars. A summary of all molecular gas detections
is given in Solomon \& Vanden Bout (2005). Some more recent CO
detections of z$\sim$2 QSOs have been presented by Coppin (2008,
Figure~16) who also discuss a possible evolutionary link between QSOs
and SMGs.

In a few cases, the quasars are sufficiently bright in CO that their
CO emission can be imaged at high spatial resolution. Perhaps the most
stunning example so far is the spatially resolved complex CO
distribution of the z=4.4 QSO BRI1335-0417 -- this source is seen when
the universe was less than 2\,Gyrs old. (Figure~17, Riechers et al.
2008). Its spatial and velocity structure of the molecular reservoir
is inconsistent with a simple gravitationally bound disk, but
resembles a merging system.  The observations are consistent with a
major, gas-rich (`wet') merger that both feeds an accreting
supermassive black hole (causing the bright quasar activity), and
fuels a massive starburst that builds up the stellar bulge in this
galaxy.  This quasar host galaxy may thus be the most direct
observational evidence that z$>$4 wet mergers at high redshift are
related to AGN activity (Riechers et al.\ 2008).

In general, spatially and kinematically resolved observations of the
molecular gas distribution offer the exciting prospect to constrain
the dynamical mass of the quasar host galaxy. This mass can then be
compared to the mass of the central black hole. An open question is
still whether the tight relation between central black hole mass and
the surrounding stellar spheroid in nearby massive ellipticals (e.g.,
Magorrian \& Binney 1998, Gebhardt et al.\ 2000, Ferrarese \& Merritt
2000), roughly M$_{\rm bulge}$/M$_{\rm BH}\sim$700 is also seen at
large lookback times. This would hint at a co--eval evolution of the
central black hole and the surrounding stellar bulge -- a remarkable
scenario by the way if one considers that the black hole is nine
orders of magnitudes smaller in size than the surrounding stellar
bulge. As the stellar bulge can not be directly observed in quasars at
high redshift (due to dimming and the very bright point source due to
central accretion), obtaining an upper limit through dynamical mass
measurements could be a promising way forward to constrain a possible
change in this ratio.  Current evidence so far points at a much lower
M$_{\rm bulge}$/M$_{\rm BH}$ ratio than seen locally (e.g. Walter et
al.\ 2004, Riechers et al.\ 2008b, 2009, Weiss et al.\ 2007, Coppin et
al.\ 2008, Carilli \& Wang 2006), though a clear answer will have to
await higher sensitivity and resolution observations of a sizeable
sample with ALMA.

\begin{figure}
\begin{center}
\includegraphics[height=.4\textheight,angle=0]{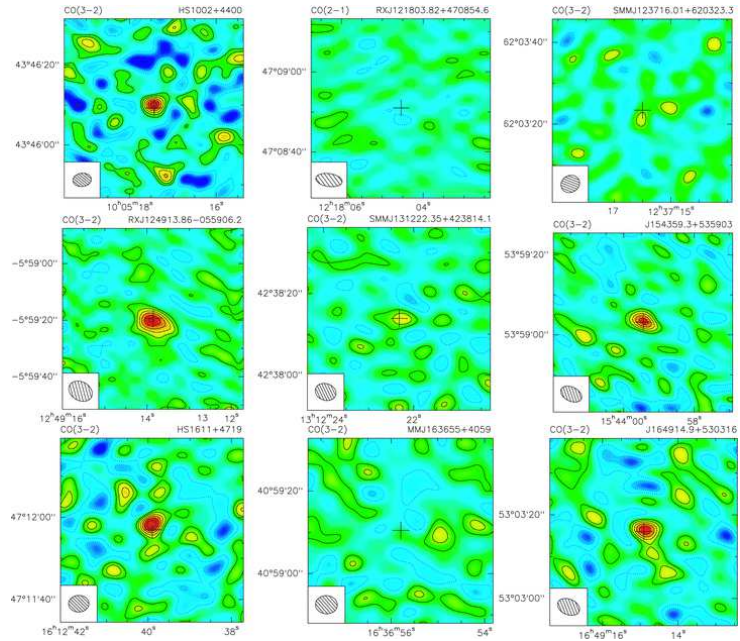}
\end{center}
\caption{Detection of molecular gas in a sample of z$\sim$2 QSOs using
  the Plateau de Bure interferometer.  Figure taken from Coppin et
  al.\ (2008).}
\end{figure}

\begin{figure}
\begin{center}
\includegraphics[height=.3\textheight,angle=0]{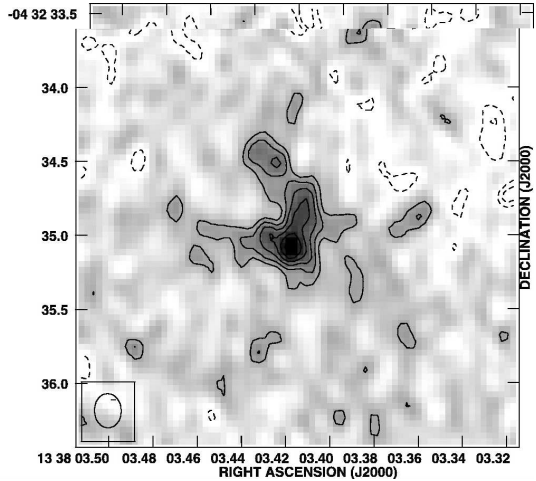}
\end{center}
\caption{Spatially resolved CO distribution in the z=4.41 QSO
  BRI1335--0417 obtained with the VLA. The high resolution (0.15$''$
  or $\sim$1\,kpc at this redshift, see beamsize in lower left
  courner) reveals a complex structure in the molecular gas emission,
  possibly hinting at a (wet) merger.  Figure taken from Riechers et
  al.\ (2008).}
\end{figure}

Like in the case of the SMGs, Spitzer IRS spectroscopy also resulted
in detection of PAH emission features in high--redshift quasars (Lutz
et al.\ 2007, 2008, Martinez-Sansigre et al.\ 2008). As shown in the
composite spectrum of z$\sim$2.5 QSOs in Figure~18, the various PAH
features are nicely detected and the PAH luminosity and rest frame
far-infrared luminosity correlate and extend a similar correlation
seen in the case of lower luminosity local QSOs (Lutz et al.\ 2008).
Like in the case of the SMGs, these measurement provide additional
strong evidence for intense star formation activity in the hosts of
these millimeter--bright QSOs.  This is yet another argument that the
rest--frame FIR luminosity in quasars is predominantly powered by star
formation, and not AGN activity.

\begin{figure}
\begin{center}
\includegraphics[height=.3\textheight,angle=0]{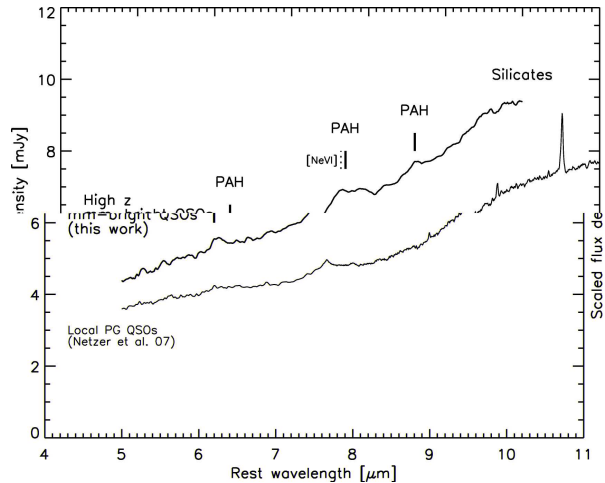}
\end{center}
\caption{Spitzer IRS composite spectrum of a dozen millimeter--bright
  quasars at redshift z$\sim$2.5. The PAH complexes are clearly
  detected, suggesting major star formation activity in these objects.
  Figure taken from Lutz et al.\ (2008).}
\end{figure}

\subsection{Prospects of dust detection in the Epoch of Reionization}

As mentioned earlier, current observations are limited to the
brightest objects in the early universe. ALMA, with its
orders--of--magnitude increase in sensitivity, will be able to detect
more typical source populations at high redshifts. It is very likely
that ALMA will even reveal a major population of dusty systems at
redshifts beyond 6 (when the universe was less than a Gyr old), as
dust production has apparently been quite effective in the very early
universe. E.g. in the case of the z=6.42 quasar J1148+5251, FIR
emission was first detected by Bertoldi et al.\ (2003a).  The SDSS
observations, Keck spectroscopy, and HST imaging (White et al.\ 2003,
2005), reveal a SMBH of 2$\times$10$^9$ M$_\odot$ and a very compact
structure. The host galaxy has now been detected in thermal dust,
non-thermal radio continuum, CO line, and [CII] 158 $\mu$m emission
(Figure~19). High resolution imaging of the CO emission reveals a
massive reservoir of molecular gas, 2$\times$10$^{10}$ M$_\odot$,
distributed on a scale of 6 kpc in the host galaxy (Walter et al.
2003, 2004). The broad band SED of J1148+5251, shows a clear FIR
excess, consistent with 50K dust emission and with the radio-FIR
correlation for star forming galaxies (Wang et al.\ 2008). The high CO
excitation in J1148+5251 (Bertoldi et al.\ 2003b, Riechers et al.,
subm.) is comparable to that seen in starburst nuclei implying dense
(10$^5$\,cm$^{-3}$), warm gas.  The high--resolution imaging of the
[CII] emission in the host galaxy of J1148+5251 reveals an extreme
starburst region with a diameter of 1.5kpc (Maiolino et al.\ 2005,
Walter et al.\ 2009), forming stars at the `Eddington limited' rate of
1000 M$_\odot$\,year$^{-1}$\,kpc$^{-2}$ (Thompson et al.\ 2005).
These measurements demonstrate that major star formation episodes are
happening in the first Gyr of the Universe and that they lead to major
dust production on galactic scales, as evidenced by the FIR and
molecular gas measurements.  Clearly, it will be critical to extend
similar dust studies to less extreme systems at similar and higher
redshifts. The key point to take away from this discussion is that
dust production appears to be very efficient (in at least a few
systems) when the Universe was less than a Gyr old (see also Maiolino
2004).

\begin{figure}
\begin{center}
\includegraphics[height=.2\textheight,angle=0]{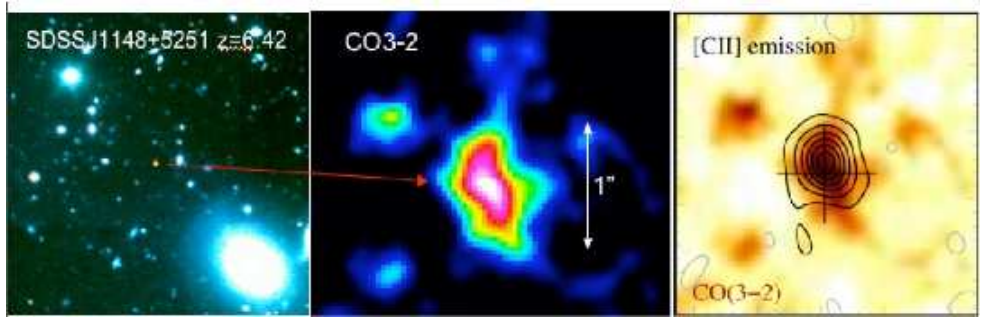}
\end{center}
\caption{Images of SDSS J1148+5251 at z = 6.42. Left is a Keck
  true-color image (Djorgovski, Mahabal, and Bogosavljevic, priv.
  comm.). Center is the VLA image of CO 3-2 emission (Walter et al.\
  2004). Right is the PdBI [CII] 158$\mu$m image (Walter et al.\
  2009).}
\end{figure}

\section{Concluding remarks}

This review can not do justice to all the work that has been done in
the field of high redshift dust emission over the last decade.  But it
hopefully provides an overview over the variety of strategies and
techniques that have been used to characterize dust emission (and thus
L$_{\rm FIR}$) in systems all the way out to the epoch of reionization
(z$>$6). The field has come a long way over the last decade ---
however it is also clear that at the same time it is still in its
infancy. The systems that are currently being studied are certainly
the `freaks' amongst the typical galaxy population at high redshifts,
as even major local starforming systems such as Arp\,220 are too faint
for detection at significant redshifts (z$>$0.5) with current
facilities. The future, however, is bright: in only a couple years
from now, the IRAM Plateau de Bure interferometer will have completed
its upgrade to full 8\,GHz bandwidth, an order of magnitude increase
in bandwidth compared to Bure's performance just a few years back. A
little later, ALMA, given its site, collecting area, receiver
technology and bandwidth, will afford another order of magnitude
increase in sensitivity, which without doubt will revolutionize this
field of research.

\acknowledgements

I thank Henrik Beuther, Chris Carilli and Rob Ivison for comments on
the manuscript. It is my pleasure to thank my colleagues Frank
Bertoldi, Chris Carilli, Pierre Cox, Emanuele Daddi, Helmut
Dannerbauer, Thomas Greve, Dominik Riechers, Ian Smail, Ran Wang and
Axel Weiss.


\end{document}